\documentstyle[12pt,epsfig]{article}

\def\beq{\begin{eqnarray}}
\def\eeq{\end{eqnarray}}
 \def\ni{\noindent}

\begin{document}
\newcommand{\rf}[1]{{\rm (\ref{#1})}}
\newcommand{\ord}[2]{{\cal O} \left( { {#1} \over {#2} } \right)}

\thispagestyle{empty}

\hfill  gr-qc/0109085

\hfill  \today

\vskip1cm \centerline{\Large \bf
 Regular
Sources of the Kerr-Schild class for } \centerline{\Large \bf
 Rotating
and Nonrotating Black Hole Solutions} \vskip.8cm
\par
\centerline{\sc Alexander Burinskii}
\smallskip
\centerline{Gravity Research Group, NSI, Russian Academy of
Sciences } \centerline{B. Tulskaya 52  Moscow 113191 Russia; \
e-mail: bur@ibrae.ac.ru} \vskip.4cm\centerline{\sc Emilio
Elizalde\footnote{Presently on leave at the Department of
Mathematics,  Massachusetts Institute of Technology, 77
Massachusetts Avenue, Cambridge, MA 02139-4307.}}
\smallskip
\centerline{Instituto de Ciencias del Espacio (CSIC) \, \& }
\centerline{Institut d'Estudis Espacials de Catalunya (IEEC/CSIC)}
\centerline{Edifici Nexus, Gran Capit\`a 2-4, 08034 Barcelona,
Spain; \  e-mail: elizalde@ieec.fcr.es}
\vskip.4cm\centerline{\sc Sergi R. Hildebrandt}
\smallskip
\centerline{Institut d'Estudis Espacials de Catalunya (IEEC/CSIC)}
\centerline{Edifici Nexus, Gran Capit\`a 2-4, 08034 Barcelona,
Spain; \  e-mail: hildebrandt@ieec.fcr.es} \vskip.4cm
\centerline{\sc Giulio Magli}
\smallskip
\centerline{Dipartimento di Matematica del Politecnico di Milano,
} \centerline{Piazza Leonardo Da Vinci 32, 20133 Milano, Italy; \
e-mail: magli@mate.polimi.it}
\vskip12mm

\centerline{\bf Abstract}
\vskip2mm

{\small  A unified approach to regular interiors of black holes with smooth
matter distributions in the core region is given. The approach is
based on a class of Kerr-Schild metrics representing minimal
deformations of the Kerr-Newman solution, and allows us to give a
common treatment for (charged and uncharged) rotating and
nonrotating black holes. It is shown that the requirement of
smoothness of the source constraints the structure of the core
region in many respects: in particular, for Schwarzschild holes a
de Sitter core can be selected, which is surrounded by a smooth
shell giving a leading contribution to the total mass of the
source. In the rotating, noncharged case the source has a similar
structure, taking the form of a (anisotropic and rotating) de
Sitter-like core surrounded by a rotating elliptic shell. The Kerr
singular ring is regularized by anisotropic matter rotating in the
equatorial plane, so that the negative sheet of the Kerr geometry
is absent. In the charged case the sources take the form of
``bags'', which can have  de Sitter or anti de Sitter interiors and
a smooth domain wall boundary, with a tangential stress providing
charge confinement. The ADM and Tolman relations are used to
calculate the total mass of the sources.
}

\vfill

\noindent {\it PACS:}  04.70.-s, 97.60.Lf, 04.20.Jb, 04.20.Dw, 04.20.-q

\par


\section{Introduction}
\label{s-intro}
This paper is an attempt at a unification of two research lines
on black hole solutions, which have been developed (almost
independently) for a long time. A first line of investigation has
to do with the problem of the final state in gravitational
collapse, and stems from pioneering observations by Gliner
and Sakharov \cite{Gl,Sak}, who suggested that matter at superhigh
densities should have the equation of state $p=-\epsilon$, so that
the the stress-energy tensor takes the ``lambda term'' form
\begin{equation}
T_{ik}=\Lambda g_{ik}
\label{TL}
\end{equation}
at the late stage of collapse. Further, Zel'dovich proposed
\cite{Zel} such a stress-energy tensor to arise as the result of
gravitational interactions in a vacuum polarization process. These
considerations led naturally to the hypothesis that an unlimited
increase of spacetime curvature during the collapse process had to
be halted by the formation of a core region with a constant,
limiting value of the curvature determined by dominant effects of
quantum fluctuations. The issue received renewed attention
over twenty years later, essentially following the papers by Frolov,
Markov and Mukhanov \cite{fmm1,FMM}. Their model consist of a de Sitter
core inside a Schwarzschild black hole, matched with the external
solution via a thin transition layer. All investigations along
this line have been restricted, so far,  to the non rotating case
only \cite{PIsr}--\cite{bor2}
(see also \cite{ayon1}--\cite{ayon3} for the Reissner-Nordstr{\"o}m case).
\par
Another open line of research is connected with the
analysis of the structure of the singularity of the rotating (Kerr
and Kerr-Newman) black holes. As is well known, this singularity
takes the form of a ring which is a branch line of the space,
leading to a two-sheeted topology.
Going through the Kerr's ring one obtains a second
(`negative') sheet of the metric where the values of the mass and
charge change their signs while fields change their directions. In this
region closed timelike curves exist, so that causality violations
occur. This led to approaches which attempted to avoid the
two-sheetedness with procedures meant to truncate the negative
sheet. A procedure of this kind was first developed by Israel
\cite{Is}, who used the surface of the disk spanned by the singular
ring as the surface of truncation. The resulting metric has a
finite jump of the first derivative on the disk, thus  leading to
a  distributional matter source located on the surface. In this
way the Kerr solution is interpreted as being the field generated
by a very exotic stress-energy tensor: a layer of
negative mass rotating with superluminal velocities!

The Israel interpretation was improved
by Hamity \cite{ham}, who noticed that
the disk can be considered as being in
rigid relativistic rotation.
In the co-rotating reference system, the
stress-energy tensor takes a diagonal form,
with zero energy density and a negative pressure,
which however grows up to
infinity on approaching the singular boundary of the disk.

Another approach was given, for the Kerr-Newman solution, by
L\'opez \cite{Lop}, who constructed the source in the form of a
rigidly rotating ellipsoidal shell (bubble) covering the
singular ring. The singular ring is removed, since the interior of
the bubble is flat. A continuous matching of the flat
interior with the external metric of the Kerr-Newman field can be
obtained, however, only by a special choice of the shell
``radius'' $r_{shell}=r_e= \frac {e^2}{2m}$, where $r$ is the Kerr
ellipsoidal radial coordinate.

Actually, $r_e$ is the so called `classical size' of a particle
with charge $e$ and mass $m$, and indeed the problem of Kerr's
sources received attention also from the point of view of
constructing classical models of the electron, after Carter's
remark \cite{Car} that the Kerr-Newman solution possesses the same
giromagnetic ratio $g=2$ of the Dirac electron.\footnote{In this
connection a series of works followed, on the
models of spinning particle based on the Kerr-Newman solution
\cite{Is,Lop,Bur0}--\cite{Bur}.} Interestingly enough, as
far as models of spinning particles are concerned, the values of
charge $e$ and the rotation parameter $a=J/m$ are very high with
respect to the mass $m$ so that the horizons of the Kerr-Newman
solution are absent and the Kerr singular ring is a naked
singularity, visible also to far-away observers.
\par

In spite of the progress in understanding the structure of regular
black hole solutions in both the aforementioned approaches, there
was still one common drawback connected with the necessity of
involving in the models a thin (or at least infinitely thin) transition
layer, while smooth models for the black hole interior would
obviously be more satisfactory
\cite{Dym,Kr,GG,I,Mag,ayon1,ayon2,ayon3,Lop1,Bur1,Bur}.
In many such attempts, the treatment is based upon the Kerr-Schild
class of metrics \cite{I,GG,Lop1,Mag,Bur}, what is obviously connected with
the fact that all stationary black holes (that is, all {\it known} black
holes) are particular cases of the Kerr--Schild geometry. On the
other hand, the de Sitter core region can be described in
Kerr-Schild form, too. Therefore, it looks convenient to describe
in this same form the transition region connecting the core
and external geometry. A remarkable property of the Kerr-Schild
class is that such a description can be performed in a {\it unique}
fashion for charged, uncharged, rotating, and nonrotating BH
solutions, by using a smooth function of a radial coordinate $f(r)$
to interpolate between the core and the external field. This is the
approach used in the present paper. Sources of BH solutions are
constructed as smooth deformations of the electro-vacuum Kerr-Schild
metrics retaining the main structure of this geometry, namely the
double principal null (PN) congruence. In this way we obtain a class of
sources which covers almost all previous models of nonrotating
sources \cite{FMM,PIsr,Dym,DymSol,EH,I,mmsen,Mag,ayon1,ayon2,ayon3}
generalizing them to the rotating case. It contains smooth analogues
of known shell-like models, in particular, the rotating and the
nonrotating shell (bubble) models of charged sources
\cite{CC,Lop,Lop2}. Several new interesting features appear in
this way.

Throughout this work, latin indices run from 0 to 3. We write Einstein's
equations in the form $ - R_{ab} + (1/2) g_{ab} R = 8 \pi T_{ab} $, where $
R_{ab} $ is the Ricci tensor, $ R_{ab} \equiv R^c_{abc} $, and  units
are chosen so that $ G=1 $, $ c=1 $. The Lorentz signature is taken to
be $-+++$.

\section{Generalized Kerr--Newman solutions}
\label{s-gkns}

The Kerr-Newman solution in the Kerr-Schild form is \cite{DKS}
\begin{equation}
g_{ik} =\eta_{ik} +2 h e^3 _i e^3 _k \ .
\label{gt}
\end{equation}
Here $\eta_{ik}=$ diag $(-1,1,1,1)$, $h$
is a scalar function (to be specified below), and
the null vector field $e^3 _i$  is tangent to the Kerr PN
congruence.
The explicit form of the Kerr PN congruence is not essential for
our analysis. We will assume that in the generalization of the
Kerr-Newman solution to the interior case, the Kerr PN congruence
retains its form and also the properties of being geodesic and shear
free.

A simple way to generalize the Kerr-Newman solution ---obtaining an
interior solution which is still of the Kerr--Schild type--- is to
replace the factor $f_{KN}=mr - e^2/2$ in the function $h$ with an
arbitrary function $f(r)$ \cite{GG}. This procedure can be seen as
the introduction of a smooth  distribution for charges and masses,
This distribution is purely ``radial'' in that it depends only on the
$r$-coordinate, which is
confocal to the angular coordinate $\theta$ of
an oblate ellipsoidal coordinate system, and becomes the standard
``tortoise'' radial coordinate if the oblateness goes to zero
(see e.g. \cite{MTW}).

The Kerr-Schild metric takes a convenient form in a basis
where the one-form $e^3$ is normalized
in such a way that its
time component equals one.
In this basis the function $h$ takes the form
\begin{equation}
h= \frac {f(r)}\Sigma = \frac{f(r)}{r^2 + a^2 \cos ^2 \theta} \ .
\label{hgt}
\end{equation}
For regular $f(r)$, this function can be singular only in the
equatorial plane $\theta= \pi/2 $, at $r=0$. The behavior near
$r=0$ is of the form $h\sim f(r)/r^2$, as in the nonrotating case
(when $a=0$). This allows us to apply the same approach to the
regularization of the metrics, both for for the nonrotating and
for the rotating cases.
\par
By using the ansatz (\ref{hgt}) and the machinery  of the
Debney--Kerr--Schild approach \cite{DKS}, we obtain  the following
tetrad components for the Ricci tensor
\begin{eqnarray}
R^{\prime}_{12}&=& - 2G, \\
R^{\prime}_{34} &=& D+ 2G, \\
R^{\prime}_{12}-R^{\prime}_{34}&=& -(D +4 G), \\
R^{\prime}_{23} &=& (D+4G) (r,_2 - P_{\bar Y }),\\
R^{\prime}_{13} &=& (D+4G) (r,_1  - P_Y )    , \\
R^{\prime}_{33} &=& - 2 (D+4G) (r,_1  - P_Y)(r,_2  - P_{\bar Y}) ,
\label{Rprabt}
\end{eqnarray}
where
\begin{equation}
D= - \frac{f^{\prime\prime}}
{\Sigma},
\label{Dt}
\end{equation}
\begin{equation}
G= \frac{f'r-f}{\Sigma^2}
\label{Gt}
\end{equation}
(here $f^{\prime}=\partial_r f(r);$ and $P_{\bar Y} = \partial_{\bar Y} P =
2^{-1/2} Y; \quad P_{Y} = \partial_{Y} P = 2^{-1/2} \bar Y$;
$,_a= e_a^{i}\partial _i$ are the tetrad derivations, the
corresponding tetrad is given in App.~\ref{app-a}, \rf{tetpr1}--\rf{tetpr4}).
This expression allows us to write the stress-energy tensor
(App.~\ref{app-a}, (\ref{fT})), which acquires a very transparent form if
a orthonormal tetrad
$\{u,l,m,n\}$ (\ref{orttetr}) connected with Boyer-Lindquist coordinates is
used (App.~\ref{app-c}):  \begin{equation} T_{ik} = (8\pi)^{-1} [(D+2G) g_{ik}
- (D+4G) (l_i l_k -  u_i u_k)].  \label{Tt} \end{equation} In the above
formula, $u^i$ is a timelike vector field given by $$ u^i=\frac
1{\sqrt{\Delta\Sigma}}(r^2+a^2,0,0,a),$$ where $\Delta= r^2 + a^2 -2 f(r)$ .

In this expression one immediately recognizes that, if the matter of the
source is thought of as being separated into ellipsoidal layers corresponding
to constant values of the coordinate $r$, each layer rotates with angular
velocity $\omega(r)= \frac {u^{\phi}}{u^0}=a/(a^2+r^2)$. This rotation becomes
rigid only in the thin shell approximation $r=r_0$. The linear velocity of the
matter w.r.t. the auxiliary Minkowski space is $v=\frac {a \sin
\theta}{\sqrt {a^2 + r^2}}$, so that on the equatorial plane
$\theta =\pi /2$, for small values of $r$ ($r\ll a $), one has $v
\approx 1$, that corresponds to an oblate, relativistically
rotating disk.

The energy density $\rho$ of the material satisfies
to $T^i_ku^k=-\rho u^i$ and is, therefore,  given by
\begin{equation}
\rho = \frac{1}{8\pi} 2G.
\label{rhot}
\end{equation}
There are only two distinct
spacelike eigenvalues, corresponding to the
radial and tangential pressures of the non rotating case, namely
\begin{equation}
p_{rad} = -\frac{1}{8\pi} 2G=-\rho,
\label{pradt}
\end{equation}
\begin{equation}
p_{tan} = \frac{1}{8\pi}(D+ 2G)=\rho +\frac{D}{8\pi}. \label{prtt}
\end{equation}
Singularities can arise only at $r=0$ on the equatorial plane, so
that the regularity properties of the stress-energy tensor can be
studied together in both the rotating and the nonrotating cases
(energy conditions will be dealt with in Sect.~\ref{sec-ecs}).

\section{Interiors for the Kerr--Schild class of BH solutions}
\subsection{The non-rotating case}

We shall consider in this section the nonrotating case from a unifying
viewpoint, before moving to the rotating case.\subsubsection{Core region}
\par\noindent We follow here a somewhat classical
approach, motivated by well known (and expected) properties of the
strong field regime of gravitation (see e.g. \cite{FMM}). We thus
consider the center of the core region as a spacetime of constant
curvature. Therefore, we have to use for this region a
regularizing function $f(r)=f_0(r)= \alpha r^4,$ where $\alpha
=\frac {8\pi\Lambda}{6}$. This provides smoothness of the metric
up to the second derivative and removes the singularity at $r=0$.
The scalar curvature invariant $R=2D= - 2 f_0^{\prime\prime}/r^2 =
- 24\alpha$ is constant, as well as the density $\rho =
\frac{1}{8\pi}\Lambda $.\subsubsection{Exterior region}\par\noindent
As is obvious, the exterior is determined unambiguously by the
Birkhoff-Israel theorem. Thus the function $f$ must coincide
with $f_{KN} = mr -e^2/2$.
\subsubsection{Transition region}
\par\noindent We assume a De-Sitter like behaviour of the spacetime only near
the center (``core'' region). Therefore, between the boundary
of the source - assumed to lie in the region of trapped surfaces -
and the de Sitter core, a `transition region' can exist,
which interpolates between the core and the vacuum in such a way that the
resulting metric is regular everywhere.
Due to the assumed simmetries, different kinds of transitions
correspond to different choices of the
function $f(r)$.  Calculating the second fundamental form of the $r=$
const surfaces, it is easy to check that,
to avoid the presence of singular (shell-like)
distributions of matter on the inner (DeSitter-transition) and on
the outer (transition-vacuum) matching hypersurfaces,
the function $f$ must be $ C^1 $.
Further, one can ask the transition region to be `thin' in the sense
that the thickness $\delta$ is much smaller than the radial position of
the transitional layer $r_0$ which would correspond to a cusp at the
intersection of the plots of $f_0(r)$ and $f_{KN} (r)$.
As a result of these assumptions the position of the transition layer
can be easily estimated analytically. It is, however, more transparent the
use of a graphical representation of the different kinds of situations
which may occur.


\par\noindent
As a result of these assumptions the position of the transition layer
can be easily estimated as a root $r_0$ of the equation
$f_0(r_0) =f_{KN}(r_0)$.
We shall see that this relation turns out a necessary condition for
consistency of the source models with respect to the Tolman and
ADS mass relations. It is, however, more transparent the use of a
graphical representation.

\subsection{Graphical analysis}
\label{ss-graph}\subsubsection{Case $\alpha >0$: de Sitter
interior, uncharged source}
\begin{figure}[ht]
\centerline{\epsfig{figure=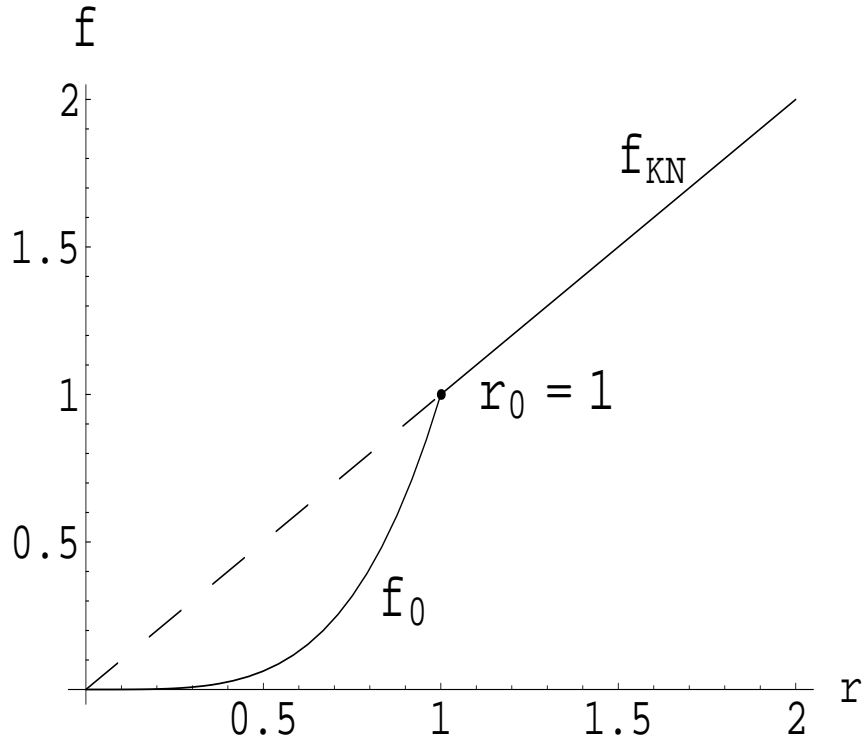,height=10cm,width=12cm}}
\caption{\small Position of phase transition
$ r_0 $ as an intersection of plots $ f_0 (r) $ and $ f_{KN} (r) $. Uncharged source, $ \alpha > 0 $, arbitrary units.}
\end{figure}
\par\noindent Fig. 1 shows that there is
only one intersection between $f_0(r)=\alpha r^4$ and
$f_{KN}(r)=mr$. Therefore, the position of the transition layer
will be $r_0 =(m/\alpha)^{-1/3}$. As  seen in the picture,
the second derivative of the corresponding interpolating function
will be negative at this point, yielding an extra contribution to
the positive tangential pressure in the transition region.
Solutions of this class can be constructed in such a way that the
weak energy condition is satisfied (see Sect.\ref{sec-ecs}).
\subsubsection{Case $\alpha >0$: de Sitter interior, charged
source}
\begin{figure}[ht]
\centerline{\epsfig{figure=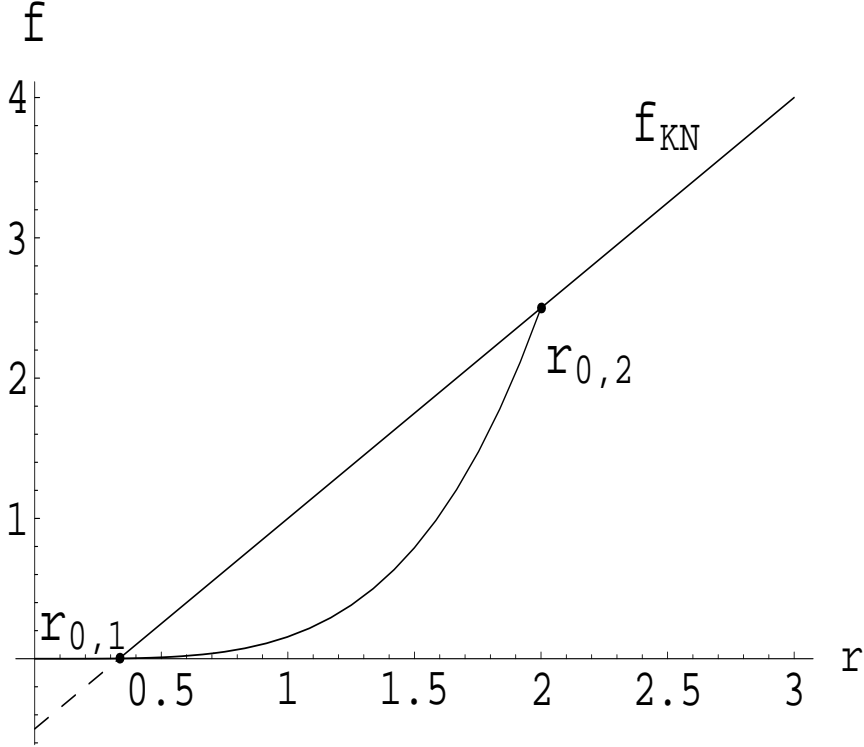,height=10cm,width=12cm}}
\caption{\small Two possible positions for point of phase transition $ r_{0,1} $ and $ r_{0,2} $ for a charged source and $ \alpha > 0 $.}
\end{figure}
\par\noindent Fig. 2 shows that for charged sources there
are two intersections, $r_{0,1}$ and $r_{0,2}$, of the functions
$f_0(r)=\alpha r^4$ and $f_{\rm KN}(r)=mr-e^2/2$. The root
$r_{0,2}$ corresponds to a negative second derivative of the
function $f(r)$ and leads to a picture similar to that for the uncharged
source with the intermediate shell with pressure (Fig. 3).
\begin{figure}[ht]
\centerline{\epsfig{figure=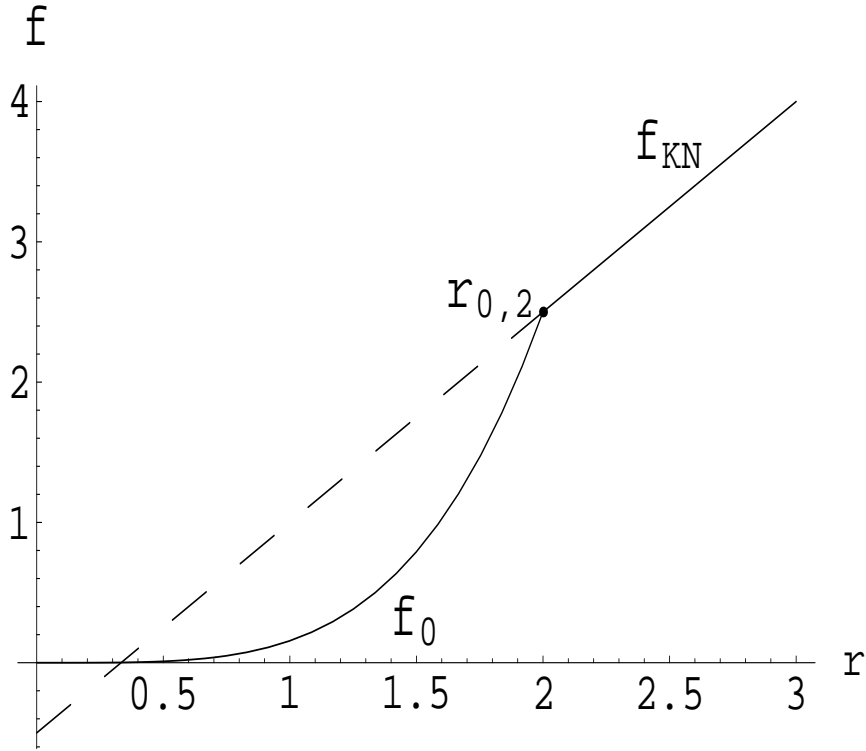,height=10cm,width=12cm}}
\caption{\small Choice of $ r_{0,2} $ as point of phase transition.}
\end{figure}
However, the smaller root $r_{0,1}$ corresponds to a source of
smaller size and has a positive second derivative leading to an
intermediate shell with a tangential stress. Thus, this source
resembles a bubble with a de Sitter interior and a domain wall
boundary confining the charge of the source (Fig. 4).
\begin{figure}[ht]
\centerline{\epsfig{figure=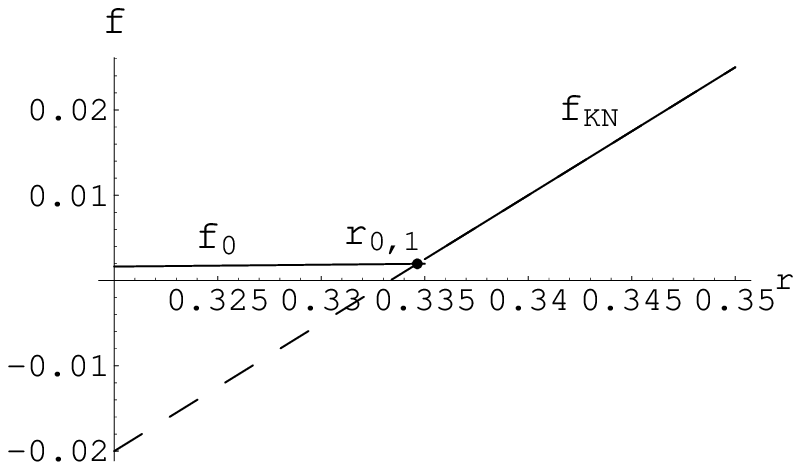,height=10cm,width=12cm}}
\caption{\small Strongly scaled section of Fig. 1 corresponding to
phase transition at point $ r_{0,1} $.}
\end{figure}
 \subsubsection{Case $\alpha < 0$: anti-de Sitter interior, uncharged source}
\begin{figure}[ht]
\centerline{\epsfig{figure=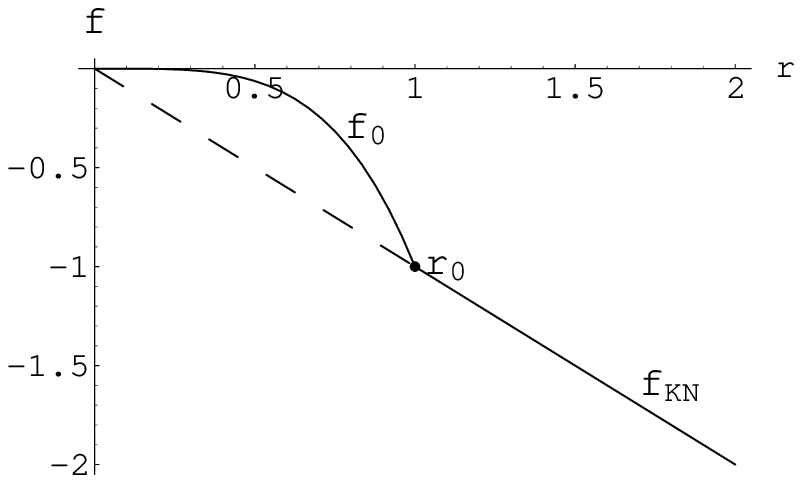,height=10cm,width=12cm}}
\caption{\small Nonstable case corresponding to $ \alpha < 0 $ and negative mass.}
\end{figure}
\par\noindent In this case,  matching the interior and exterior regions
turns out to be  impossible, for the usual black hole solutions.
However,  there is an exotic case of a black
hole solution with negative mass. Graphical analysis shows (Fig. 5)
that there is a solution with positive  second derivative of the
interpolating function $f(r)$ leading to a shell  with tangential
stress. This exotic source resembles an AdS bubble with domain
wall boundary and negative total mass, like those occurring in
supergravity \cite{CvS}.\subsubsection{Case $\alpha \le 0$:
flat and anti-de Sitter interior, charged source}
\begin{figure}[ht]
\centerline{\epsfig{figure=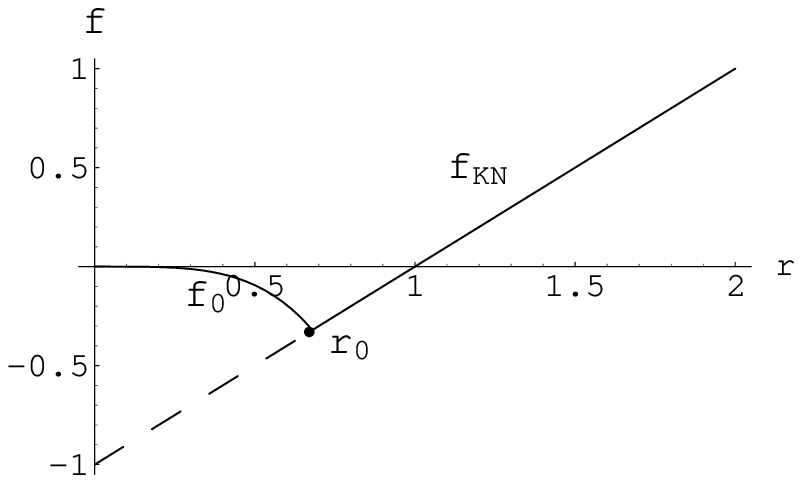,height=10cm,width=12cm}}
\caption{\small Class of stable states for charged sources and $ \alpha \le 0 $ (flat or AdS interior and $ r_0 \le e^2/2m $).}
\end{figure}
\par\noindent
In this case a matching of the interior and the exterior regions can be
reached via an intermediate shell, having a positive second derivative
and leading to a positive tangential stress of the shell confining
the charge of the source.

This graphical analysis provides a classification of the
simplest possible matching of AdS and dS interiors of the source
with the exterior black hole solutions via a smooth intermediate
layer. Of course,  more complicated transition layers can be considered,
where the second derivative of the interpolating function $ f(r) $
changes sign several times.

\subsection{The rotating case}
A remarkable property of the Kerr--Schild class is that
the above treatment is easily extensible to the rotating case. In
fact, the function $f$ is, also in this case, a function of $r$
only, and the Kerr and Kerr-Newman solutions are obtained with the
{\it same } functions that correspond to the Schwarzschild and to the
Reissner-Nordstr\"om solutions, respectively. Of course, the
definition of the coordinate $r$ is now completely different, since
the surfaces $r=$ const are ellipsoidal, and described by the
equation
\begin{equation}
\frac{x^2 +y^2}{r^2 + a^2} + \frac {z^2}{r^2} =1.
\label{ell}
\end{equation}
The relations for the metric and stress-energy tensor are
characterized by a more complicated form of the function $\Sigma = r^2
+ a^2 \cos ^2 \theta$. As a result, the components of the metric and
stress-energy tensor increase when approaching the equatorial plane
$ \cos \theta = 0$, where they take the same form as in the
nonrotating case. The singularity of the Kerr solution can be
suppressed, by analyzing the metric and the stress-energy tensor near
the Kerr singular ring $r \approx \cos \theta \approx 0$. As we
have already seen, the condition on the behavior of the function $f$ when
$r\to 0$ remains the same as for the nonrotating case.

The intermediate shell, which matches the interior of the source
and the external geometry, is foliated on the rotating ellipsoidal
layers. The thin shell is characterized by an increasing of the
tangential stress (or pressure) and can be considered as a rigidly
rotating boundary of the disk-like source, similarly as in the
L\'opez singular shell model \cite{Lop}.

\section{Causal structure}
\label{s-cs}

The causal structure of nonrotating black hole interiors is well
known\cite{bor1,mmsen}. It closely resembles that of the RN spacetime,
with the key difference that the singularity is replaced by the
matter-filled region. From the topological point of view, it was
shown by Borde that a change of topology occurs, making it
possible for regular solutions to exist\cite{bor1,bor2}.

The case of a rotating black hole interior will now be analyzed.
This is best visualized, again, with the use of a two-dimensional
plot in the equatorial plane. Horizons of the spacetime, if any,
are defined by $\Delta=0$, that is, by the equation
$f(r)=r^2+a^2$. Plotting the parabola and taking into account the
properties of the function $f$, one realizes that two different
situations may occur.\par\noindent {\bf Case I}: the
parabola $ r^2 + a^2 $ and the external
function $f$ intersect at one point, which
corresponds to a new Cauchy horizon replacing the vacuum one (the
event horizon obviously remains in the vacuum
region).\par\noindent {\bf Case II}: For large enough values of
the angular momentum the intersection between the two previous plots
 is absent, displaying the disappearance of both the Cauchy and event horizon.

\section{Local energy conditions and interpretation}
\label{sec-ecs}

The physical reasonability of a classical solution of the Einstein
equations relies on the fulfillness of the energy conditions,
which are a common tool for getting insight into some energetic
properties of any spacetime. However, and in spite of the fact
that a quantum theory of gravity is not available yet, increasing
evidence, both experimental and theoretical (see e.g.
\cite{casimir}--\cite{hajicek}, the recent account in
\cite{visser} and references therein) shows that quantum effects
---specially those associated with the quantum vacuum--- may lead to
a general violation of some or even {\em all} of such conditions.
Therefore, as we aim at describing nonsingular black holes which
are only likely to be understood via quantum effects, we should
not expect all the energy conditions to be satisfied (actually,
the strong and the dominant energy conditions have to be violated
somewhere, already on  classical grounds \cite{mmsen}). In any
case, energy conditions are a useful tool which may still serve to
assess whether a classically dominated field is the source
responsible for the  spacetime considered, or, on the contrary,
that such a possibility is banned and one should seek for a
quantum origin of the source. This remark is important because,
otherwise, energy conditions might impel us to disregard some
spacetimes that may well fit with the current knowledge of the
interface between quantum physics and gravitation. Let us begin
now with the study of {\em local} energy conditions. In the
following section we will implicitly consider some {\em averaged}
---or ``extended''--- energy conditions.

Let $ {\vec V} $ be any 4-velocity vector field and let $ \vec N $
be any null vector field. We shall denote $ S_{\lambda \mu}
T^{\lambda} T^{\mu} $ by $ S_{TT} $, for any (symmetric) rank-two
tensor field, $ \bf S $, and any vector field $ \vec T $.
\smallskip

\ni {\sf (i) Strong energy condition (SEC):} a system satisfies
the SEC iff \beq - R_{VV} \propto T_{VV} + (1/2) T \geq 0, \quad
\forall \vec V , \eeq where $ R_{\lambda \mu} $ is the Ricci
tensor and $ T_{\lambda \mu} $ is the stress-energy tensor (being
$ T $ its trace). Fulfillment of SEC is a ``cornerstone" in
singularity theorems and must be violated in the case the
spacetime is nonsingular. It can be shown, that also the dominant
energy condition ($T^\mu_\nu V^\nu$ non-spacelike for any
non-spacelike $V$) is violated in this case.
\smallskip

\ni {\sf (ii) Weak energy condition (WEC):} a system satisfies the
WEC iff \beq T_{VV}  \geq 0, \quad \forall \vec V . \eeq These are
clearly connected with the sign of the energy density measured by
an observer with 4-velocity $ \vec V $.
\smallskip

\ni {\sf (iii) Null energy condition (NEC):} a system satisfies
NEC iff \beq T_{NN} \geq 0, \quad \forall \vec N . \eeq This last
energy condition may be viewed as a limiting case of (ii) for
ultra relativistic observers. Basically, it includes some commonly
used spacetimes, as anti de Sitter.

The physical validity of (i) has been objected many times and on
different grounds, and violation of the SEC is nowadays well
understood. On the other hand, (ii) is usually considered as a
necessary condition for a gravitational system to be acceptable.
While this is certainly right for classical matter, it becomes
doubtful in the case when quantum effects play a relevant role.

It is useful to represent $ \vec V $  as
\beq
{\vec V }= A {\vec u} + B {\vec l} + C {\vec m} + E {\vec n} \nonumber
\eeq
where $ \{ {\vec u}, {\vec l}, {\vec m}, {\vec n} \} $ is any basis of the
tangential space. In particular we will choose the {\em orthonormal} basis
given in App.~\ref{app-c}, which corresponds to a comoving observer, because
it gives raise to simpler expressions in this case.
Since $ {\vec V} $ is timelike one has
$$ A^2 = 1 + B^2 + C^2 + E^2 . $$
From~\rf{Tt} one can get
$$ 8 \pi T_{VV} = - (D+2G) - (D+4G) (B^2-A^2) = 2 G + (D+4G) (C^2 + E^2),
$$ and
$$ 8 \pi T  = 2 D, $$
whence
$$ T_{VV} + (1/2)T = (D+2G) + (D+4G) (C^2+E^2). $$
We can get a more direct expression, since a direct computation shows that
\beq
\label{eq-cons}
D + 4 G = - {\Sigma \over r} G' .
\eeq
Finally, this can be written in terms of the pressure, stress and energy
density measured by the comoving observer using
$$ 8 \pi \rho = 2 G, \quad 8 \pi p_{rad} = - \rho, \quad 8 \pi p_{tan} =
(D + 2G). $$
The result is
\beq
T_{VV} =\rho  - {\Sigma \over 2 r } \rho' (C^2 +E^2), \quad T_{VV} + {1
\over 2} T = p_{tan} - {\Sigma \over 2 r } \rho' (C^2 +E^2),
\eeq
where $\rho' =\partial _r \rho.$
It follows that:
\par

\ni (i) The SEC is satisfied iff $ p_{tan} \geq 0 $ and $ \rho'
\leq 0 $, in the region of study.

\ni (ii) The WEC is satisfied iff $ \rho \geq 0 $ and $ \rho' \leq
0 $, in the region of study.

Notice that $ \rho' $ plays an essential role in both cases,
besides the more natural quantities $ p_{tan} $ and $ \rho $. We
can now analyze what happens in each region. The exterior field
has $ f_{KN}(r) = mr - e^2/2 $. Therefore, $ \rho = p_{tan} = e^2/
\Sigma^2 $. Whence, one has $ \rho, p_{tan} > 0 $ and $ \rho' < 0
$. SEC and WEC are thus obviously satisfied. For the rest of the
analysis, it is worth considering first the nonrotating case,
where $ \rho = 6 \alpha $, $ p_{tan} = - 6 \alpha $ and $ \rho' =
0$. Therefore, for $ \alpha > 0 $ (de Sitter) WEC is satisfied
whereas SEC is not and the singularity is avoided. For $ \alpha <
0 $ (anti de Sitter) WEC is clearly violated and the singularity
is again avoided.

As long as we aim at describing general properties, it is worth
keeping the freedom of choice of  $ f(r) $ in the shell region.
The main conclusion is that only for a de Sitter core ($\alpha>0$)
plus an exterior electromagnetic field satisfying $ {6 \alpha
r_1^4  \over e^2} > 1 $ (which includes the uncharged case) may
the WEC be satisfied  throughout the whole system. The reason is
as follows: if $ {6 \alpha r_1^4  \over e^2} < 1 $ then $ \rho_0
$($= 6 \alpha $) is smaller than $ \rho_{ext} $ ($= \rho_{KN} =
{e^2 \over r_1^4} $) at $ r_1 $. Therefore, the function $ f(r) $
in the transition region ---which is at least $ C^1 $ in that
region--- must be increasing in some open interval.

In fact the restriction $ {6 \alpha r_1^4 / e^2} > 1 $ is
fulfilled in most previous attempts
\cite{FMM,PIsr,BalPo,Dym,DymSol,mmsen,I}. Surprisingly, in the
models \cite{ayon2,ayon3}, one can easily see that the WEC and SEC
are found to be violated. For this it suffices to compute the
density $ \rho $ in these models. In the models of \cite{ayon2}
one has $ f(r) = [m r^4 / (r^2+ e^2)^{3/2}] \exp{(-e^2/2mr)} $ and
therefore (recall that these examples are non-rotating solutions)
$ 8 \pi \rho = 2 G = 2(f' r - f)/r^4 = [e^2/r(r^2+e^2)^{3/2}]
[6mr/(r^2+e^2)+1] \exp{(-e^2/2mr)}] $. For the models in
\cite{ayon3} one has $ f(r) = mr [ 1 - \tanh{(e^2/2mr)} ] $ and $
8 \pi \rho = e^2/[r^4 \cosh^2{(e^2/2mr)}] $. Now, in both cases
one has $ \rho(r \to 0^{+}) \to 0^{+} $ and $ \rho(r \to +\infty)
\to 0^{+} $. Consequently, there are open regions where $ \rho $
is an increasing function and therefore SEC and WEC are violated.
This adds new examples of SEC and/or WEC violations (see e.g.
\cite{visser} for a recent review), due in this case to non-linear
electrodynamics, and also is a warning about the actual relevance
of their fulfillment.

Finally, let us consider the rotating case.
In the core one has $ \rho = 6 \alpha r^4/ \Sigma^2 $, $ p_{tan} =
- 6 \alpha r^2 (r^2 + 2 a^2 \cos^2 \theta)/\Sigma^2 $ and $ \rho'
= 12 \alpha a^2 r^3 \cos^2 \theta / \Sigma^3 $. For $ \alpha > 0
$, it is clear that both SEC and WEC (because $ \rho' > 0 $) are
violated. For $ \alpha < 0 $, it is clear that WEC is violated,
although the SEC is satisfied.

The main conclusion is that the WEC and SEC are ---unavoidably---
violated in the rotating case, except at the equatorial plane,
which follows the pattern of the non-rotating case, already
explained above. For the case of NEC, similar computations bring
to the conclusion that the it is fulfilled iff $ \rho'< 0 $ and
therefore, previous considerations show that it is again
generically {\em violated} inside the object.

At first sight, this might be considered as a drawback of these
models. However, the whole thing is saying {\em only} that the
models cannot account for a {\it classical} interior of a
Kerr-Newman spacetime, so that the nature of the source should be
sought within quantum field theory. But this, in turn, is often
considered to be the natural framework to work in (e.g. assuming
interior models suggested by supergravity or string theory).

\section{Contributions to the total mass
coming from different regions of the regularized sources}
\label{s-tolman}
There are basically two ways to study the energy contributions from
different parts of the source.
 One consists in evaluating the contributions
coming from both the energy density and from the pressures of the
system, whereas the other takes into account the contribution of
the energy density only. The first is called the {\em Tolman mass}
of the system and the latter, the {\em ADM mass}. In order for
these quantities to be well-defined it is necessary that the
system be stationary and asymptotically flat (see e.g.
\cite{LL,MTW}). Our models do fulfill such requirement. These
masses are computed for an observer at rest with respect to the
asymptotically flat region. The tetrad corresponding to such
observer is explicitly given in Apps.~\ref{app-a} and~\ref{app-b},
in terms of the Kerr tetrad forms and the Kerr coordinates.

Given the previous conditions, the Tolman mass is defined by
\beq
M_{\rm Tol}=\int_{\Omega} dx^3\, \sqrt {-g}{(T_1^1 +T_2^2 + T_3^3 - T_0^0)}
\label{Tolm}
\eeq
(numerical values of the indexes refer to
asymptotically Cartesian components).
It explicitly takes into account the
contributions to the gravitational mass coming from the energy
density and the pressure of the matter forming the source
\cite{Tol,LL} ($ \Omega $ is a region of  the source). Now, using
the expressions for the Kerr tetrad forms and the stress energy
tensor given in Apps.~\ref{app-a} and~\ref{app-b}, we get \beq
M_{\rm Tol}=(8\pi)^{-1}\int_0^{\pi} \, d\theta \int_0^{2\pi} \, d\phi
\int_{r_a}^{r_b} \, dr 2 [ D+2G + (D + 4G) a^2\sin^2 \theta
/\Sigma ] \Sigma \sin \theta . \label{Tol-a} \eeq On the other
hand, and given the aforementioned conditions, the ADM mass is
defined by \beq M_{\rm ADM} = - \int_{\Omega} \, dx^3 \, \sqrt{-g}
\, T^0_0 \eeq and can be expressed as follows \beq \label{Adm-a}
M_{\rm ADM} = (8 \pi)^{-1} \int_0^{\pi} \, d\theta \int_0^{2\pi}
\, d\phi \int_{r_a}^{r_b} \, dr \, [2G + (D+4G) a^2 \sin^2\theta /
\Sigma] \Sigma \sin \theta . \eeq

Another way of writting these quantities is
\beq M_{\rm Tol} = - (4
\pi)^{-1} \int_{\Omega} \, dx^3 \, \sqrt{-g} \, R_{u u} ,
\\ \hspace*{-2cm} M_{\rm ADM} = \int_{\Omega} \, dx^3 \, \sqrt{-g} \,
T_{uu},
\eeq where $  u $ is the timelike vector of the Kerr tetrad, i.e.
$ u = e^{4'} - (1/2) e^{3'} $ using the notation of
Apps.~\ref{app-a} and~\ref{app-b}. The corresponding observer is
at rest with respect to the asymptotically flat region. From these
expressions, one sees that the positivity of the Tolman mass is
related with an averaged form of the strong energy condition,
while the ADM mass is (obviously) connected with the weak energy
condition. As we shall see, the fact that our models are
nonsingular and therefore that the strong energy condition must be
violated (at least close to the core) will be reflected in the
negative values of the Tolman mass in some regions.

We will describe the main properties of the sources coming from an
analysis of the Tolman and ADM masses for each of the three
regions of the model: interior (core), transition (shell) and
exterior region. We will also estimate their relative
contributions.

\subsection{Non-charged, non-rotating sources}

It is instructive to consider first the simplest case of
regularized sources for nonrotating, neutral black hole solutions.
In this case the treatment is  very transparent and, moreover, it
exhibits the main peculiarities common to all the models. In
particular, there is an unexpectedly large contribution to the
total mass coming from the thin shell on the boundary of the
source.

Let us start with the Tolman mass. Setting $a=e=0$ in the
relation~\rf{Tol-a}, we get
\beq
M=(8\pi)^{-1}\int \, d\theta \int \, d\phi \int \, dr \,
2(D+2G) r^2  \sin \theta
\label{Tol-0}
\eeq
which, in terms of the function $f(r)$, takes the form
\beq
(8\pi)^{-1}
\int_0^{\pi} \, d\theta \int_0^{2 \pi} \, d\phi \int_{r_a}^{r_b} \, dr
\, 2 (2f^\prime r -2f -r^2 f^{\prime\prime}) r^{-2} \sin \theta =
\nonumber \\
\int_{r_a}^{r_b} (2f^\prime r -2f -r^2f^{\prime\prime})r^{-2}dr =
[2(f / r) - f']_{r_a}^{r_b}.
\label{M-0}
\end{eqnarray}
In this case the model is composed of three regions \beq f(r) =
\cases{f_{\rm int}=\alpha r^4, & $ 0 \le r_0 $ \cr f_{\rm
shell}(r), & $ r_0 \le r \le r_0(1 +\delta) $ \cr f_{\rm ext} = m
r, & $ r_0(1+\delta) \le r $.} \eeq For the interior region we
obtain
\begin{equation}
M_{\rm int} = - [2 \alpha r^3]_0^{r_0} = -2\alpha r_0^3.
\end{equation}
Consider the case of positive $\alpha$ (de Sitter core) and
assume the shell to be thin, i.e. $ \delta << 1 $, the position
of the shell is determined, to first order in $ \delta $, by
equation $m r_0\approx \alpha r_0 ^4$, i.e. $m\approx\alpha
r_0^3$, and consequently
\begin{equation}
M_{int} \approx -2m.
\label{Mint0}
\end{equation}
Therefore, we obtain a remarkable result: the de Sitter core gives
a {\it negative } contribution $-2m$ to the total mass of the source.
Since the total mass is determined by the parameter $m$, this
means that the shell on the boundary of the core has to give a
contribution $+3m$ to the total mass. Indeed, calculating
(\ref{Tol-0}) for the region with $f(r)=f_{\rm shell}$, and
assuming again the shell to be thin, we obtain
\begin{equation}
M_{\rm shell} =
- f_{ext}^\prime \vert _ {r=r_0(1+\delta)} +
f_{int}^{\prime}\vert_{r=r_0} + {\cal O} (\delta) \approx 3 m,
\label{Ms0}
\end{equation}
providing the balance of the total mass  $M_{\rm int} + M_{\rm
shell} =m $ (the exterior contribution being zero in this case).
One should note, that here the shell is assumed to be sufficiently
thin but not necessarily infinitely thin. Besides, the exact form
of the function $f_{\rm shell}$, and consequently the matter
distribution on the shell, are not essential, giving a
contribution of order $\delta$. Irrespectively of the value of $
\delta $, the contribution from the internal de Sitter core is
always negative. We could say that the energetic properties of the
core account for the avoidance of the singularity.
\par
Let us now compare the result with the ADM mass ($a = e = 0$)
\begin{equation}
M_{\rm ADM} = (8 \pi )^{-1} \int_0^{\pi} \, d\theta \int_0^{2\pi} \, d\phi
\int_{r_a}^{r_b} \, dr 2 G r^2 \sin \theta = [f/r]_{r_a}^{r_b} .
\label{ADM}
\end{equation}
It is clear that the ADM mass does not take into account the
pressure components of the stress-energy tensor, and consequently,
it ``does not feel'' the shell region, i.e. $ M_{\rm ADM, shell} =
{\cal O}(\delta) $. However, the ADM mass obviously yields the
right value for the total mass of the source $M_{\rm ADM, int \, +
\, shell} = m$. The solution of this apparent contradiction can be
found in that, at least for the regularized sources with a thin
shell, the contribution of the pressure  components to the total
mass can be viewed as representing a gravitational ``dipole''
$(-3m,+3m)$ which forms the bubble.

\subsection{General case. Charged and rotating source.}

In the general charged and/or rotating case, it is worth writing
all the expressions in terms of the function $ f $. For the Tolman
mass, we get \beq \label{TM} M_{\rm Tol} = \int_{0}^{1}\, dx
\int_{y_a}^{y_b} \, dy \, {2(y { f^\prime _y} - f)(y^2+2-x^2)- {
f^{\prime \prime}_{y}} ( x^2 + y^2 ) ( 1 + y^2 ) \over a (x^2 +
y^2)^2 }, \eeq where we have put  $ x \equiv \cos \theta $ and $ y
\equiv r/ a $ ($ {f'_y} \equiv df(y)/dy $, $ f''_{y} \equiv
df'_y/dy $). This change makes the integrand dimensionless and
allows one to have control on the limiting cases $ a << r $ or $
a >>r $.
After some integrations, we get the remarkable result that \beq
M_{\rm Tol}  = \left\{ {f(r) \over r} + {(a^2+r^2) \over a r}
\arctan{(a/r)} [f(r)/r - f'(r)] \right\}_{y_a}^{y_b}. \eeq Hence,
the Tolman mass of a layer of the source may be obtained as the
difference between the boundary values of a suitable {\em
potential function} (the function between curly brackets above)
for any fixed value of the angular momentum (including of course
the nonrotating case).\footnote{From this result it is easy to see
that our matching condition $f(r) \in C^1$ turns out to be
necessary and sufficient for the consistency of the Tolman mass.}

Let us now consider the contributions to the Tolman mass coming
from each region (in the sequel we omit the tag ``Tol'' in the
masses). We have \beq f(r) = \cases{\alpha r^4, & $ 0 \le r_0 $
\cr f_{\rm shell}(r), & $ r_0 \le r \le r_1 $ \cr m r - e^2/2, & $
r_1 \le r $.} \eeq For the core region, we get \beq M_{\rm core} =
\alpha r_0^3 \left[1 - 3 \left({a \over r_0} + {r_0 \over a}
\right) \arctan \biggl({a \over r_0}\biggr) \right]. \eeq In the
limit $ a / r_0 << 1 $, e.g. low rotating black holes and fixed $
r_0 $, including the non-rotating limit, one gets \beq M_{\rm
core} = \cases{-2\alpha r_0^3 \left[ 1+ (a^2 / r_0^2) - (a^4 / 5
r_0^4 ) + \ord{a^6}{r_0^6} \right], \cr -2 \alpha r_0^3, & $ a= 0.
$ } \eeq In the limit $ r_0 / a << 1 $, e.g. rapidly rotating
black holes or very small core ``radius'' $r_0$ with respect to
the size of the removed singular ring ---a typical case for
parameters of spinning particles--- we have \beq M_{\rm core} =
-(3 \pi \alpha r_0^2 a / 2) \left[ 1 - (8 r_0 / 3 \pi a) + (r_0^2
/ a^2) + \ord{r_0^3}{a^3}  \right]. \eeq As in the case considered
previously, for a de Sitter core the contribution of the core
itself to the Tolman mass of the object is negative, for any value
of $ r_0 $ and $ a $, and satisfies $ M_{\rm core} \le - 2 \alpha
r_0^3 $.

For an anti de Sitter core the situation is opposite, that is, $
M_{\rm core} \ge 2 |\alpha| r_0^3 $.

The Tolman mass of the exterior region comes from the
electromagnetic field and is \beq \label{tol-em} M_{\rm exterior} =
{e^2 \over 2 r_1} \left[ {\displaystyle 1+ \biggl( {r_1 \over a} +
{a \over r_1} \biggr) \arctan \biggl({a \over r_1} \biggr) }
\right]. \eeq

Now the limits $ a << r_1 $ (including the nonrotating case) and $
r_1 << a $, yield \beq M_{\rm exterior} = \cases{(e^2 / r_1) \left[
1 + (a^2 / 3 r_1^2) - (a^4 / 15 r_1^4) + \ord{a^6}{r_1^6} \right],
& $ a << r_1 $ \cr (e^2 / r_1), & $ a = 0 $  \cr (\pi e^2 a / 4
r_1^2) \biggr) \left[ 1 + (r_1^2 / a^2) - (4 r_1^3 / 3 \pi a^3) +
\ord{r_1^5}{a^5}  \right], & $ r_1 << a $.\cr} \eeq We now have to
consider the contribution from the shell.

\subsubsection{Thin shell}

The first situation we will deal with is the  thin shell approach.
The thickness of the shell $\delta$ must satisfy the condition $
\delta / r_0 \equiv r_1 /r_0 - 1 <<1 $ (but not necessarily $
\delta = 0 $). In this case only the continuity of the
interpolating function $ f(r) $ is necessary, when dismissing all
terms of order $ \delta $ or higher. The (only) matching condition
is then \beq f_{\rm core} (r_0) = f_{\rm exterior} (r_0).
\label{r0} \eeq This approach allows us to recover from our models
the previous ones involving the assumption of singular
distributions both for the stationary and for the static cases.

The Tolman mass is, in this case,
\beq
\begin{array}{rcl}
\displaystyle
M_{\rm thin\, shell} & = & \displaystyle -[f'_0] \left[1 + (r_0 ^2/ a^2)
\right] \int_0^1 { dx \over x^2 + (r_0/a)^2} \cr
& = & \displaystyle -[f'_0] \biggl( {a \over r_0} +
{r_0 \over a} \biggr) \arctan \biggl( { a \over r_0} \biggr),
\end{array}
\eeq where $ [f'_0] \equiv f_{\rm exterior}' (r_0) - f'_{\rm core}
(r_0) = m - 4 \alpha r_0^3 = -3 m + 2 e^2/ r_0 $ and we have
used the matching condition \rf{r0}. First, we note that the {\em
total} Tolman mass of the model is $m $, as expected. That is, ($
\chi \equiv e^2/2m r_0 $), \beq
\begin{array}{l}
M_{\rm core} + M_{\rm thin\, shell} + M_{\rm exterior}  = \hfill \cr
m (1 - \chi) \left[1 - 3 \left({a \over r_0}
+ {r_0 \over a} \right) \arctan \biggl({a \over r_0}\biggr) \right] \cr
  + m ( 3 - 4 \chi ) \biggl( {a \over r_0}  +
{r_0 \over a} \biggr) \arctan \biggl( { a \over r_0} \biggr)
\allowbreak \cr + m \chi \left[ { 1+ \biggl( {r_1 \over a}
+ {a \over r_1} \biggr) \arctan \biggl({a \over r_1} \biggr) } \right] \cr
\hfill=  m .
\end{array}
\eeq

We next consider the relative contribution of each part. These
will clearly depend on the ratio $ \chi $.
\subsubsection{Astrophysical sources}
 One can see that for the case of astrophysical (neutral
or weakly charged) sources, where $ \chi = 0 $ or $ \chi << 1$ the
thin shell gives the major contribution to the total mass $ 1\le
|M_{\rm thin\, shell} / M_{\rm core} | \le 3/2 $, for any value of
$ r_0 $ and $ a $. In this case the core gives a negative
contribution to the Tolman mass, whereas the shell yields a
positive one ---bigger in absolute value than the core mass.
\subsubsection{ Strongly charged sources: particle-like solutions}
Let us first note that, classically, the electromagnetic mass of a
charged sphere of radius $r_0$ is given by $M_{\rm cl-em} =\frac
{e^2}{2r_0}$. Therefore, the parameter $\chi $ can be expressed as
$\chi=M_{\rm cl-em}/m$. The case $ \chi \sim 1$ corresponds to
strongly charged sources, where an essential part of the mass is
thought to be of electromagnetic origin, as for example in
classical models of the electron. The relation between the core
and the shell depends on the value of $ \chi $ in this
case.\footnote{This situation occurs also in the case of
\cite{ayon1,ayon2,ayon3}, in which solutions for nonrotating
regularized black holes are given.} In particular, for $\chi =1$
the core does not yield any contribution to the mass. This result
comes from \rf{r0}, since $\alpha r_0^3 = m-M_{\rm cl-em} =0$.
Therefore, $\alpha=0$ and the core is flat in this case. On the
other hand, the thin shell contribution to the mass is negative
\begin{equation}
M_{\rm thin \, shell} =
  - m  \biggl( {a \over r_0}  + {r_0 \over a} \biggr)
\arctan \biggl( { a \over r_0} \biggr).
\end{equation}
For nonrotating sources, setting $a=0$ one obtains $ M_{\rm thin\,
shell} = -m $ and $M_{\rm exterior} =2m$. In this case one can
show that  $M_{\rm exterior}$ splits into a pure electromagnetic
contribution $M_{\rm cl-em}=m$ and a gravitational contribution to
the mass coming from the electromagnetic field $M_{\rm
grav-em}=m$, and providing the balance $ M_{\rm thin\,shell}+
M_{\rm cl em}+ M_{\rm grav-em}= m$.\footnote{Similarly to the
models in \cite{Lop,Lop2}.} In the limit of a singular shell, this
case corresponds to the classical model of a charged particle
considered by Cohen and Cohen, \cite{CC}, which is a modification
of the known Dirac classical model for an extended electron
\cite{Dir}.
\par
For $ \chi > 1 $, it follows from (\ref{r0}) that $\alpha r_0^3 =
m-M_{\rm cl em}=m (1-\chi)$, and consequently, $\alpha<0$. Thus,
there must be an anti de Sitter space in the core. Graphical
analysis immediately shows that in this case the characteristic
radius $r_0$ is smaller than the classical one. The relation
between the contributions of the core and of the shell is found to
be $ 4/3 \le |M_{\rm thin\, shell} / M_{\rm core}| \le 2 $, where
now the core yields a positive contribution to the total Tolman
mass, and the shell a negative one. Notice that, except for the
case of $ \chi = 3/4 $, the Tolman mass of the object undergoes a
sudden change when passing the shell. This is because, except for
that case, $ M_{\rm thin\, shell} \neq 0 $ and $ \delta \approx 0
$.
\par\noindent
\subsection{Comparison with the ADM expression for the total mass}
\par\noindent
>From expressions \rf{Adm-a}, one can see that, contrary to what
happens in the non-rotating case, the term $D$ does give a
contribution, due to a Lorentz effect associated with the rotation
of the source. Performing the integrations in analogy with
previous calculations, one can obtain the result that the ADM mass
may also be interpreted as coming from a {\em potential function}.
The expression is \beq \label{admpot} M_{\rm ADM} = {1 \over 2}
\left\{ {f \over r} + f' + {(a^2+r^2) \over a r} \arctan{(a/r)}
[f/r - f'] \right\}_{r_a}^{r_b} . \eeq Therefore, the ADM mass of
the core region is \beq M_{\rm ADM, \, core} = \alpha r_0^3
\left[3 - \frac 12 \left({a \over r_0} + {r_0 \over a} \right)
\arctan \biggl({a \over r_0}\biggr) \right]. \eeq The ADM mass for
the exterior region comes from the pure electromagnetic part and
is  twice smaller than the Tolman one (since it does not take into
account the gravitational contribution of the electromagnetic
pressure). The expression is therefore \beq M^{\rm ADM}_{\rm
exterior} = {e^2 \over 4r_1} \left[ {\displaystyle 1+ \biggl( {r_1
\over a} + {a \over r_1} \biggr) \arctan \biggl({a \over r_1}
\biggr) } \right]. \eeq Finally, the ADM contribution from the
thin shell is \beq M_{\rm ADM, \, thin\, shell} = ([f'_0]/2)
\left[ {\displaystyle -1- \biggl( {r_1 \over a} + {a \over r_1}
\biggr) \arctan \biggl({a \over r_1} \biggr) } \right]. \eeq One
can see that in the case when the position of the thin shell is
determined by the matching condition \rf{r0}, the balance of the
total ADM mass is also attained (the general case being also
fulfilled from expression~\rf{admpot} and $ f \in C^1$), i.e.
$$M^{\rm ADM}_{\rm core}+ M^{\rm ADM}_{\rm thin\, shell}+ M^{\rm
ADM}_{\rm ext} = m. $$

Finally, let us notice that the ADM and Tolman masses are related through a
simple expression, namely, \beq 2 M_{\rm ADM} - M_{\rm Tol} =
[f'(r)]_{r_a}^{r_b} . \eeq

\section{Conclusions}

We have here considered a  wide class of smooth sources for
black holes, which includes virtually all the models considered
previously in the literature and extends them to the rotating
case, thus obtaining a unified framework for arbitrary values of
the charge and angular momentum. For nonrotating BH solutions the
sources contain a core region representing a spacetime with a
constant curvature ($\Lambda$ term) and a thin (but finite)
transitional region (spherical shell) connecting the source with
an external black hole geometry. In the case of rotating BHs this
shell acquires an ellipsoidal form (a disk of radius $\sqrt { a^2
+ r_0^2} $ and thickness $ r_0$). For a thin shell, the rotation
can be considered as rigid, with angular velocity given by $
\omega = a /(a^2 + r_0^2)$. For large angular momentum, $a\gg
r_0$, the rotation is relativistic and disklike sources are highly
oblate. The curvature is not strictly constant in the interior of
the disk and it is concentrated in a tube-like neighborhood of the
former Kerr singular ring near the border of the disk (the
expressions for the curvature and stress-energy tensor for this
general case are given in Sect. 2, and a complete description is
given in the Appendices). This class of sources includes previous
models like \cite{FMM,PIsr,Dym,I,ayon1,ayon2,ayon3} generalizing them to the
rotating case, and contains the smooth analogs of known shell-like
rotating and nonrotating models as well. In particular, for a
special choice of parameters, $r_0= \frac {e^2}{2m}$, the model
acquires a flat interior and turns into the L\'opez model of the
Kerr-Newman source \cite{Lop}. However, we have shown that, in the
general case, the bubble interior can have both a positive or a
negative scalar curvature. The matching condition \rf{r0} permits
to connect the parameters of the sources $r_0 $, $ e $, $ m $ and
$\alpha= \Lambda/6$, and to select the sources which are
consistent with respect to the mass-energy balance. Indeed, the
balance relation $\alpha r_0^3 + \frac{e^2}{2r_0}= m$ shows that
uncharged black hole solutions have positive $\alpha$,
corresponding to the de Sitter interior of the core region, while
otherwise for charged, ``small'' black holes with $
\frac{e^2}{2r_0}> m$, the sources acquire a negative $\alpha$,
yielding an anti de Sitter spacetime in the core. Such anti de
Sitter regions are also predicted for strong gravitational fields
in supergravity.

Analysis and comparison of the Tolman and ADM mass relations
allows one to determine contributions to the total mass going from
diverse regions of the source. Both expressions give the correct
result at spacelike infinity. However, the ADM expression
$M_{ADM}$ does not take into account the gravitational
contribution to the total mass coming from the strong tension (or
pressure) of the thin transition shell. This contribution can be
estimated by using the Tolman relation and shows a new,
interesting feature of these models. Indeed, we prove that the
contribution to the total mass coming from the thin shell can be
extremely large, but it is anyway canceled by the contribution
from the core. Therefore, it represents a very strong
gravitational ``polarization'' of the spacetime in the form of a
bubble with a sharp and very strong boundary. This phenomenon
could very likely have observational manifestations.

In spite of the successful description here presented of virtually
all the non-rotating models and of their extension to rotating
(i.e. Kerr-Newman) models, the Kerr-Schild class of metrics might
turn out to be too restrictive in order to be able to describe
some self-consistent field models. In particular, the Casimir
effect for a superdense state in the core can be essential
\cite{casimir,eli2,Bur1}. To construct a field matter model
leading to a tangential stress of the shell, scalar fields can be
involved \cite{Bur} in the formation of an object similar to a domain wall
boundary of the bubble. However, the Kerr-Schild class (in four
dimensions) is hardly compatible with simple models of classical
scalar fields, what can be seen from the relation $T_{ik}^{KS}
e^{3i}e^{3k} = T^{KS}_{44}= 0 $ (which is one of the conditions in
the derivation  of the Kerr-Schild class of metrics \cite{DKS}.)
This argument can also be expressed in terms of quantum
corrections and the conformal anomaly for scalar fields \cite{I}.
This means that either the Kerr-Schild class of the (BH interior)
source models has to be modified, to take into account these and
other features of the ``desired'' source, or this has to be done
with the field model. In particular, the following close
generalizations can be suggested for future work in this field:
(i) conformal Kerr-Schild metrics (one of whose representatives is
the Nariai solution \cite{Nar}); (ii) inclusion of dilaton and
axion fields; (iii) field models in supergravity \cite{Bur} and
low energy string theory \cite{Sen}; and (iv) extension to higher
dimensions, in particular based on the AdS/CFT correspondence
(\cite{bur3}, see also the recent review \cite{Odin}).
\par

\vspace{9mm}

\noindent{\bf Acknowledgments}
\smallskip

One of us (A.B.) would like to thank G. Alekseev for stimulating
discussion.
This investigation has been supported by DGI/SGPI (Spain), project
BFM2000-0810, by CIRIT (Generalitat de Catalunya),
contract 1999SGR-00257, and by the program INFN (Italy)--DGICYT (Spain).
\medskip

After this paper was finished, we were informed that the advantages of the
Kerr coordinates in the analysis of rotating black holes had also been
considered in the papers by J. Ibanez, P. Papadopoulos and J. Font
\cite{Font}, and that a close problem was investigated in the paper by
A. DeBenedictis at al. \cite{DBen}. We are thankful to J. Font and
A. DeBenedictis for bringing these works to our attention and for their
comments.

\appendix

\section{Stress-energy tensor for the generalized Kerr-Schild metrics}
\label{app-a}
Starting from the following general form of the function $h$
\begin{equation}
h= f(r) /(\Sigma P^2),
\label{h}
\end{equation}
and the Kerr-Schild null tetrad
\begin{eqnarray}
e^1 &=& d\zeta  -Y dv, \\
e^2 &=& d\bar \zeta - \bar Y dv \\
e^3 &=& du + \bar Y d\zeta + Y d\bar \zeta -Y \bar Y  dv\\
e^4 &=& dv + h e^3 ,
\label{tetr}
\end{eqnarray}
we obtain, using the machinery of the Kerr-Schild formalism
\cite{DKS}, the following tetrad components of the Ricci
tensor\footnote{These calculations are very tedious and use the
extra relations given in App. B.}
\begin{eqnarray}
R_{12}&=& - 2G \\
R_{34} &=& D+ 2G\\
R_{12}-R_{34}&=& -(D +4 G)\\
R_{23} &=& (D+4G) r,_2/P \\
R_{13} &=& (D+4G) r,_1/P \\
R_{33} &=& -2 r,_1 r,_2 (D+4G) /P^2,
\label{Rab}
\end{eqnarray}
where
\begin{equation}
D= - f^{\prime\prime}/(r^2 + a^2 \cos ^2 \theta),
\label{D}
\end{equation}
\begin{equation}
G= (f'r-f)/ (r^2 + a^2 \cos ^2 \theta)^2.
\label{G}
\end{equation}
This expression (\ref{Rab}) can be transformed into another, more convenient
tetrad (Eqs.~6.1 of \cite{DKS}) $e^{a \prime}$, which is connected with the
Kerr angular coordinates $(r,t,\theta,\phi)$ determined by the relations
\begin{eqnarray}
x+iy &=& (r+ia) e^{i\phi} \sin \theta, \\
z    &=& r \cos \theta, \\
\rho &=& r+ t .
\end{eqnarray}
The reverse of the (6.1 DKS) relations are
\begin{eqnarray}
e^1 &=& e^{1\prime} + P_{\bar Y} e^{3\prime}, \\
e^2 &=& e^{2\prime} + P_{ Y} e^{3\prime}, \\
e^3 &=& P e^{3\prime},\\
e^4 &=& P^{-1} [e^{4\prime} - P_Y e^{1\prime} - P_{\bar Y } e^{2\prime} -
e^{3\prime} P_Y P_{\bar Y}].
\end{eqnarray}
They yield a new expression for the Ricci tensor
\begin{eqnarray}
R^{\prime}_{12} &=& R_{12}, \\
R^{\prime}_{34} &=& R_{34}, \\
R^{\prime}_{23} &=&  P R_{23} +  P_{\bar Y} (R_{12} - R_{34}),\\
R^{\prime}_{13} &=&  P R_{13} +  P_{Y} (R_{12} - R_{34}),\\
R^{\prime}_{33} &=& P^2 R_{33} + 2 P ( P_Y R_{23} + P_{\bar Y} R_{13})
+ 2 P_Y P_{\bar Y} (R_{12} - R_{34})
\label{Rprimeab}
\end{eqnarray}
As a result,
\begin{eqnarray}
R^{\prime}_{12}&=& - 2G, \\
R^{\prime}_{34} &=& D+ 2G, \\
R^{\prime}_{12}-R{\prime}_{34}&=& -(D +4 G), \\
R^{\prime}_{23} &=& (D+4G) (r,_2 - P_{\bar Y }),\\
R^{\prime}_{13} &=& (D+4G) (r,_1  - P_Y )    , \\
R^{\prime}_{33} &=& - 2 (D+4G) (r,_1  - P_Y)(r,_2  - P_{\bar Y}) .
\label{Rprab}
\end{eqnarray}
The primed tetrad (Eqs.6.5-6.6 of the DKS-paper) takes the form
\begin{eqnarray}
\label{tetpr1}e^{1\prime}&=&
2^{-1/2}e^{i\phi}(r+ia\cos \theta)(d\theta +i\sin \theta d\phi), \\
e^{2\prime}&=&
2^{-1/2}e^{-i\phi}(r-i a\cos \theta)(d\theta - i\sin \theta d \phi), \\
e^{3\prime}&=& dr + dt -a \sin ^2 \theta d\phi, \\
e^{4\prime}&=& dr - a \sin ^2 \theta d\phi + \frac{1}{2}(2h - 1) e^{3\prime}
\label{tetpr4}
\end{eqnarray}
Dropping the primes,  we obtain the following expressions for the
energy-momentum tensor
\begin{equation}
8\pi T_{ab}=- R_{ab}+{1\over 2}g_{ab}R,
\label{Tab}
\end{equation}
\begin{equation}
8\pi T = 8\pi T_{ab}e^a e^b = (D+2G) g_{ab}e^a e^b - 2(D+4G) e^3 \tilde e^4,
\label{fT}
\end{equation}
where
\begin{eqnarray}
\tilde e^4 = e^4 +(r,_1 -P_Y) e^1 +(r,_2 -P_{\bar Y}) e^2
- (r,_1 -P_Y)(r,_2 -P_{\bar Y}) e^3 =
\nonumber \\
{1\over 2}[dr- (dt-a\sin ^2\theta d\phi)] +
\frac {2f- a^2 \sin ^2 \theta}{2 \Sigma }e^3.
\label{te4}
\end{eqnarray}

This expression for the energy-momentum tensor coincides with the
result obtained by G\"urses and G\"ursey.
\section{Tetrad forms and representation of the Kerr-Schild
class of metrics in the Kerr and Boyer-Lindquist angular coordinates}
\label{app-b}
The Kerr-Schild class of metrics has the form
\begin{equation}
g_{ik} =\eta_{ik} +2 h k _i k _k,
\label{g}
\end{equation}
where $\eta_{ik}=$ diag $(-1,1,1,1)$ is the metric of an auxiliary
Minkowski space with Cartesian coordinates $t,x,y,z$. and $ h
=\frac {f(r)}{r^2+a^2 \cos ^2 \theta}$. The Kerr angular
coordinates $(r,t,\theta,\phi)$ are determined by the relations
\begin{eqnarray}
x+iy &=& (r+ia) e^{i\phi} \sin \theta, \\
z    &=& r \cos \theta, \\
\rho &=& r+ t .
\label{Kc}
\end{eqnarray}
In these coordinates, the metric tensor has the form
\begin{equation}
g_{(Kerr) ik} =
\left( \begin{array}{cccc}
2h-1&2h&0&-2ha\sin ^2 \theta \\
2h&1+2h&0&-(1+2h)a\sin ^2 \theta \\
0&0&\Sigma&0 \\
-2ha\sin ^2 \theta&-(1+2h)a \sin ^2 \theta&0&(r^2+a^2 +2ha^2\sin ^2
\theta )\sin ^2 \theta
\end{array} \right),
\label{gK}
\end{equation}
where $\Sigma=r^2+a^2\cos ^2 \theta$ . The determinant is det
$g_{Kerr}= -\Sigma ^2 \sin ^2 \theta $, and the contravariant form
of the metric is
\begin{equation}
g_{(Kerr)}^{ik} =
\left( \begin{array}{cccc}
-(1+2h)&2h&0&0 \\
2h&\Delta /\Sigma&0&a/\Sigma \\
0&0&1/{\Sigma}&0 \\
0&a/{\Sigma}&0&(\Sigma \sin ^2 \theta) ^{-1}
\end{array} \right),
\label{gcontK}
\end{equation}
where $\Delta = r^2 +a^2 -2f(r).$
\par
The Kerr null tetrad (Eqs.6.5-6.6 of DKS-paper) has the form
\begin{eqnarray}
e^{1\prime}&=&
2^{-1/2}e^{i\phi}(r+ia\cos \theta)(d\theta +i\sin \theta d\phi), \\
e^{2\prime}&=&
2^{-1/2}e^{-i\phi}(r-i a\cos \theta)(d\theta - i\sin \theta d \phi), \\
e^{3\prime}&=& dr + dt -a \sin ^2 \theta d\phi, \\
e^{4\prime}&=& dr - a \sin ^2 \theta d\phi + \frac{1}{2}(2h - 1)
e^{3\prime}. \label{Kntetr}
\end{eqnarray}
The
contravariant components are
\begin{eqnarray}
e^{1\prime i}&=&\frac {e^{i\phi}}{\sqrt{2}(r-ia\cos \theta)}
(0,ia\sin \theta,1, 1/\sin \theta),\\
e^{2\prime i}&=&\frac {e^{-i\phi}}{\sqrt{2}(r+ia\cos \theta)}
(0,-ia\sin \theta,1, 1/\sin \theta),\\
e^{3\prime i}&=&(-1,1,0,0),\\
e^{4\prime i}&=&\frac 12(1+2h,1-2h,0,0).
\label{Kcontr}
\end{eqnarray}

One can see that the expression for the stress-energy tensor is
simplified by the introduction of the null vector $\tilde e^4 =
e^{4\prime} -Ce^{1\prime} -\bar C e^{2\prime} - C\bar C
e^{3\prime}$, where $C = r_{,1} -P_Y$. The vector $\tilde e^4$
belongs to the null tetrad obtained from $e^{a\prime}$ by a ``null
rotation'' \cite{DKS} leaving $e^3$ unchanged $\tilde e^3 =
e^{3\prime}$. The corresponding null tetrad is completed as
follows: $\tilde e^1 = e^{1\prime} +\bar C e^{3\prime}$ and
$\tilde e^2 = e^{2\prime} + C e^{3\prime}$. This tetrad is
connected with the Boyer-Lindquist representation of the Kerr
geometry, in which a symmetry between the null vectors $\tilde
e^3$ and $\tilde e^4$ appears.
\section{Kerr-Schild  metrics in Boyer-Lindquist coordinates}
\label{app-c}
Boyer-Lindquist coordinates
$\bar t, r,\theta, \bar \phi$ are connected with the Kerr angular
coordinates $t,r,\theta, \phi$
by the relations $dt=d\bar t +(2f/\Delta)dr$ and
$d\phi=d\bar \phi +(a/\Delta)dr,$
where $\Delta = r^2 +a^2 -2f(r).$
In Boyer-Lindquist coordinates, the tetrad $\tilde e^a$ takes the form
\begin{eqnarray}
\tilde e^{1}&=&
2^{-1/2}e^{i\phi}(r+ia\cos \theta)(d\theta +i\sin \theta
\frac {r^2+a^2}{\Sigma} d\bar \phi
-\frac{ia\sin \theta}{\Sigma} d\bar t), \\
\tilde e^{2}&=&
2^{-1/2}e^{-i\phi}(r-ia\cos \theta)(d\theta -i\sin \theta
\frac {r^2+a^2}{\Sigma} d\bar \phi+\frac{ia\sin \theta}{\Sigma} d\bar t), \\
\tilde e^{3}&=&\frac{\Sigma}{\Delta}dr+(d\bar t -a\sin ^2 \theta
d\bar\phi),\\
\tilde e^{4}&=&\frac {\Delta}{2\Sigma} \lbrack
\frac{\Sigma}{\Delta}dr-(d\bar t -a\sin ^2 \theta d\bar\phi) \rbrack.
\label{BLntetr}
\end{eqnarray}
In Boyer-Lindquist coordinates, the metric tensor takes the form
(bars are omitted everywhere)
\begin{equation}
g_{(BL) ik} =
\left( \begin{array}{cccc}
2f/\Sigma-1&0&0&-2af\sin ^2 \theta /\Sigma \\
0&\Sigma/\Delta&0&0\\
0&0&\Sigma&0 \\
-2af\sin ^2 \theta/\Sigma&0&0&(r^2+a^2 +
\frac{2fa^2\sin ^2 \theta}{\Sigma})\sin ^2 \theta
\end{array} \right).
\label{gBL}
\end{equation}
The determinant is det $g_{BL}= -\Sigma ^2 \sin ^2 \theta $.

The contravariant form of the metric is
\begin{equation}
g^{(BL) ik} =
\left( \begin{array}{cccc}
-\frac {r^2 +a^2 +(2f/\Sigma)a^2\sin ^2 \theta}{\Delta} &0&0&
-\frac{2af}{\Sigma\Delta}\\
0&\Delta /\Sigma&0&0 \\
0&0&1/\Sigma&0 \\
-\frac{2af}{\Sigma\Delta}&0&0&\frac {1-2f/\Sigma}{\Delta \sin ^2 \theta}
\end{array} \right),
\label{gcontBL}
\end{equation}
The orthonormal tetrad, in the Boyer-Lindquist coordinates has the
form
\begin{eqnarray}
u&=&-\sqrt{\frac{\Delta}{\Sigma}}(dt-a\sin^2 \theta d\phi), \\
l&=&\sqrt{\frac{\Sigma}{\Delta}}dr, \\
n&=&\sqrt{\Sigma}d\theta, \\
m&=&\frac{\sin \theta}{ \sqrt{\Sigma}}\lbrack adt-(r^2+a^2) d\phi \rbrack,
\label{orttetr}
\end{eqnarray}
where $u$ is the unit timelike vector and $m$ the radial one.
The corresponding contravariant components are
\begin{eqnarray}
\label{u-vec}u^i&=&\frac 1{\sqrt{\Delta\Sigma}}(r^2+a^2,0,0,a), \\
l^i&=&\sqrt{\frac{\Delta}{\Sigma}}(0,1,0,0), \\
n^i&=&\frac 1{\sqrt{\Sigma}}(0,0,1,0), \\
m^i&=&\frac {-1}{\sqrt{\Sigma}\sin \theta}(a \sin ^2 \theta,0,0,1).
\label{contrort}
\end{eqnarray}

The null vector forms $\tilde e^3$ and $\tilde e^4$ can be
expressed via $u$ and $l$ as follows $\tilde e^3 = \sqrt
{\frac{\Sigma}{\Delta}}(l-u) $, and $\tilde e^4 =\frac
12\sqrt{\frac{\Delta}{\Sigma}}(l+u) $.
\subsection{Some useful relations}
The following relations are useful for the transition from the
Kerr to the BL coordinate system
$$r,_2 - P_{\bar Y} = - ia  (\cos \theta) ,_2 =
\frac{iae^{i\phi}\sin \theta}{\sqrt 2 (r-ia \cos \theta)}, $$
$$r,_1- P_Y =  ia  (\cos \theta) ,_1 =
\frac{-iae^{-i\phi}\sin \theta}{\sqrt 2 (r+ia \cos \theta)}, $$
$$(r,_1 - P_Y )e^1 + (r,_2 - P_{\bar Y})e^2 = a\sin^2 \theta d\phi ,$$
$$(r,_1 - P_Y )(r,_2 - P_{\bar Y} )=
a^2 \sin^2 \theta /(2\Sigma). $$

\end{document}